\documentclass[aps,pra,twocolumn,amsmath,amssymb,nofootinbib,showpacs,superscriptaddress]{revtex4-1}
\usepackage[english]{babel}
\usepackage{latexsym}
\usepackage{graphics}
\usepackage{graphicx}
\usepackage{epsfig}
\usepackage{color}
\usepackage{bm}
\usepackage{amsmath}
\usepackage{amssymb}
\usepackage{amsthm}
\usepackage{dcolumn}
\usepackage{bm}
\usepackage{float}
\usepackage{hyperref}
\usepackage{color}
\usepackage{epstopdf}
\usepackage{braket}
\usepackage{cleveref}
\usepackage[svgnames]{xcolor}
\hypersetup{hidelinks,colorlinks=true,allcolors=DarkBlue}

\newcommand{\X}{\textsf{X}}
\newcommand{\RZ}{\textsf{R}_Z}
\newcommand{\RY}{\textsf{R}_Y}
\newcommand{\RX}{\textsf{R}_X}

\newcommand{\CNOT}{\textsf{CNOT\,}}

\theoremstyle{remark}

\begin{document}

\preprint{APS/123-QED}

\title{Multiclass classification using quantum convolutional neural networks \\ with hybrid quantum-classical learning}

\author{Denis Bokhan}
\email{denisbokhan@mail.ru}
\affiliation{Laboratory of Molecular Beams, Physical Chemistry Division, Department of Chemistry, Lomonosov Moscow State University, Moscow 119991, Russia}
\affiliation{Russian Quantum Center, Skolkovo, Moscow 143025, Russia}

\author{Alena S. Mastiukova}
\affiliation{Russian Quantum Center, Skolkovo, Moscow 143025, Russia}
\affiliation{National University of Science and Technology ``MISIS'', Moscow 119049, Russia}

\author{Aleksey S. Boev}
\affiliation{Russian Quantum Center, Skolkovo, Moscow 143025, Russia}

\author{Dmitrii N. Trubnikov}
\affiliation{Laboratory of Molecular Beams, Physical Chemistry Division, Department of Chemistry, Lomonosov Moscow State University, Moscow 119991, Russia}

\author{Aleksey K. Fedorov}
\email{akf@rqc.ru}
\affiliation{Russian Quantum Center, Skolkovo, Moscow 143025, Russia}
\affiliation{National University of Science and Technology ``MISIS'', Moscow 119049, Russia}

\date{\today}
\begin{abstract}
Multiclass classification is of great interest for various applications, for example, it is a common task in computer vision, where one needs to categorize an image into three or more classes.
Here we propose a quantum machine learning approach based on quantum convolutional neural networks for solving the multiclass classification problem. 
The corresponding learning procedure is implemented via TensorFlowQuantum as a hybrid quantum-classical (variational) model, 
where quantum output results are fed to the softmax activation function with the subsequent minimization of the cross entropy loss via optimizing the parameters of the quantum circuit. 
Our conceptional improvements here include a new model for a quantum perceptron and an optimized structure of the quantum circuit.
We use the proposed approach to solve a 4-class classification problem for the case of the MNIST dataset using eight qubits for data encoding and four ancilla qubits;
previous results have been obtained for 3-class classification problems.
Our results show that accuracies of our solution are similar to classical convolutional neural networks with comparable numbers of trainable parameters.
We expect that our finding provide a new step towards the use of quantum neural networks for solving relevant problems in the NISQ era and beyond. 
\end{abstract}

\maketitle

\section{Introduction}

Quantum computing is now widely considered as a new paradigm for solving computational problems, 
which are believed to be intractable for classical computing devices~\cite{Fedorov2022,OBrien2010,Monroe1998,Shor1994,Lloyd1996}. 
The idea behind quantum computing is to use quantum physics phenomena~\cite{OBrien2010}, such as superposition and entanglement.
Specifically, in the quantum gate-based model, quantum algorithms are implemented as a sequence of logical operations under the qubits (quantum analogs of classical bits), 
which compose the corresponded quantum circuits terminating by qubit-selective measurements~\cite{Monroe1998}. 
Examples of the problems, for whose quantum speedups are expected to be exponential, are prime factorization~\cite{Shor1994} and simulating quantum systems~\cite{Lloyd1996}, 
for example, modelling complex molecules and chemical reactions~\cite{Aspuru-Guzik2020}.
The amount of computing power for such applications is, however, greatly exceeds resources of currently available quantum computing devices. 
For example, factoring RSA-2048 bit key requires 20 million noisy qubits~\cite{Gidney2021}, 
whereas currently available noisy intermediate-scale quantum (NISQ) devices have about 50-100 qubits~\cite{Preskill2018}. 
Quantum computing can be also considered in the context of data processing~\cite{Lloyd2014} and machine learning applications~\cite{Dunjko2018}, 
where the required resources for solving practical problems are expected to be not so high. 
Still the caveats of quantum machine learning are related to the input/output problems~\cite{Biamonte2017}: 
Although quantum algorithms can provide sizable speedups for processing data, they do not provide advantages in reading classical input data. 
The cost of reading the input then may in some cases dominate over the advantage of quantum algorithms. 
One may note that various approaches have been suggested, specifically, amplitude encoding~\cite{Salomaa2005},
but the problem of the conversion of classical data into quantum data in the general case remains open~\cite{Biamonte2017}.

The use of NISQ devices in the context of the quantum-classical (variational) model has emerged as a leading strategy for their use in the NISQ era~\cite{Babbush2021-4,Aspuru-Guzik2022}. 
In such a framework, a classical optimizer is used to train a parameterized quantum circuit~\cite{Babbush2021-4}.
This helps to address constraints of the current NISQ devices, specifically, limited numbers of qubits and noise processes limiting circuit depths.
An interesting link between quantum-classical (variational) model and architectures of artificial neural networks opens 
up prospects for the use of such an approach for machine learning problems~\cite{Lloyd2013,Dunjko2016,Melko2018,Lukin2019-3,Woerner2019,Woerner2021,Schuld2019,Schuld2019-2}. 
The workflow of variational quantum algorithms, where parameters of circuit are iteratively updated (optimized), resembles classical learning procedures~\cite{Woerner2019}. 

\begin{figure*}[]
	\includegraphics[width=0.85\linewidth]{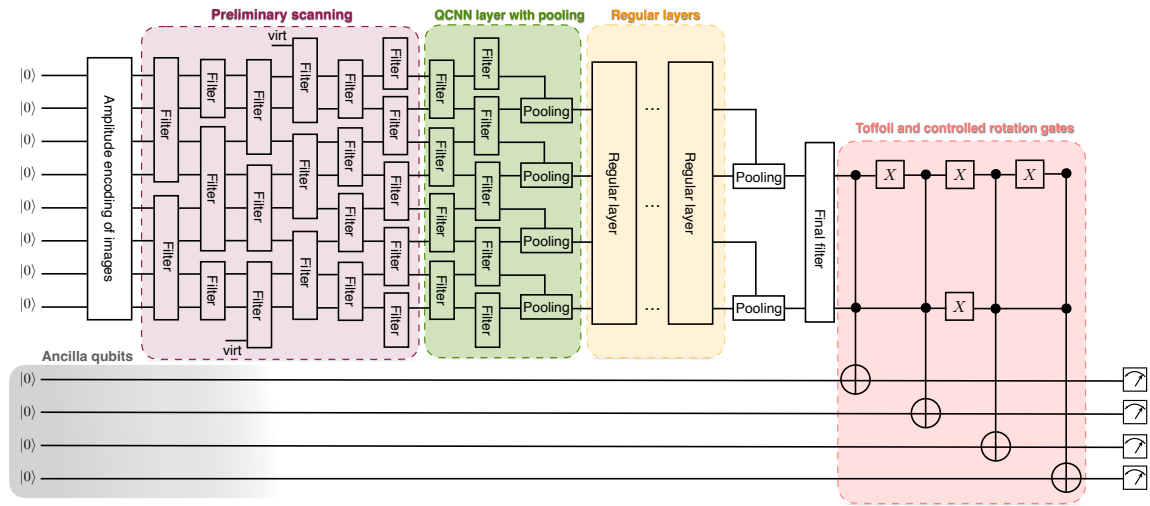}
	\vskip-4mm
	\caption{General structure of the proposed quantum neural network structure consisting of several steps: 
	Preliminary scanning using $n$-qubit filters, pooling, and regular layers.}
	\label{fig:General}
\end{figure*}

A cornerstone problem of various machine-learning-based approaches is classification, that it why it has been widely considered from the view point of potential speedups using quantum computing. 
As it has been demonstrated in Refs.~\cite{Lloyd2014,Mengoni2019}, kernel-based quantum algorithms may provide efficient solutions for the classification problem.
Specifically, the quantum version of the support vector machine~\cite{Lloyd2014} can be used as an optimized binary classifier with complexity logarithmic in the size of the vectors and the number of training examples.
A distant-based quantum binary classification has been proposed in Ref.~\cite{Schuld2017}.
Alternative versions of binary quantum classifiers have been considered in Refs.~\cite{Benedetti2018,Grant2018,Gambetta2019,Tacchino2019,Kerenidis2020} (for a review, see also Ref.~\cite{Li2021}). 
A natural next step is to consider the multiclass classification, which has been addressed recently in Ref.~\cite{Chalumuri2021} with the demonstration of the performance on the IBMQX quantum computing platform. 
This method uses single-qubit encoding and amplitude encoding with embedding of data, so the obtained results are of quite high accuracy for the 3-class classification task. 
Very recently, an approach based on quantum convolutional neural network (QCNN)~\cite{Hur2022} have been used for binary classification, albeit, 
a way to its extension to the multiclass classification case has been discussed.
We also note that some of the proposed quantum machine learning algorithms have been tested in practically relevant settings, 
for example, analyzing NMR readings~\cite{Demler2020,Demler2021} with the trapped-ion quantum computer, 
learning for the classification of lung cancer patients~\cite{Jain2020} and classifying and ranking DNA to RNA transcription factors~\cite{Lidar2018-3} using a quantum annealer, 
weather forecasting~\cite{Rigetti2021} on the basis of the superconducting quantum computer, and many others~\cite{Perdomo-Ortiz2018}. 

In this work, we present a quantum multiclass classifier that is based on the QCNN architecture. 
The developed approach use a traditional utilization of convolutional neural networks, in which few fully connected layers are placed after several convolutional layers. 
The corresponding learning procedure is implemented via TensorFlowQuantum~\cite{Babbush2020} as a hybrid quantum-classical (variational) model, 
where quantum output results are fed to softmax cost function with subsequent minimization of it via optimization of parameters of quantum circuit. 
Then we discuss the modification of a quantum perceptron, which enables us to obtain highly accurate results using quantum circuits with relatively small number of parameters. 
The obtained results demonstrate successful solving the classification problem for the 4-classes of MNIST images. 

Our paper is organized as follows.
In Sec.~\ref{sec:algorithm}, we present the general description of the proposed quantum algorithm that is used for multiclass classification.
In Sec.~\ref{sec:layers}, we provide in-detail discussion of the layer of the proposed quantum machine learning algorithm.
In Sec.~\ref{sec:results}, we demonstrate the results of the implementation of the proposed algorithm for multiclass image classification for hand-written digits from MNIST and clothes images from fashion MNIST datasets.
We conclude in Sec.~\ref{sec:conclusion}.

\section{General scheme}\label{sec:algorithm}

\begin{figure*}[ht]
        \includegraphics[width=12cm,height=6cm]{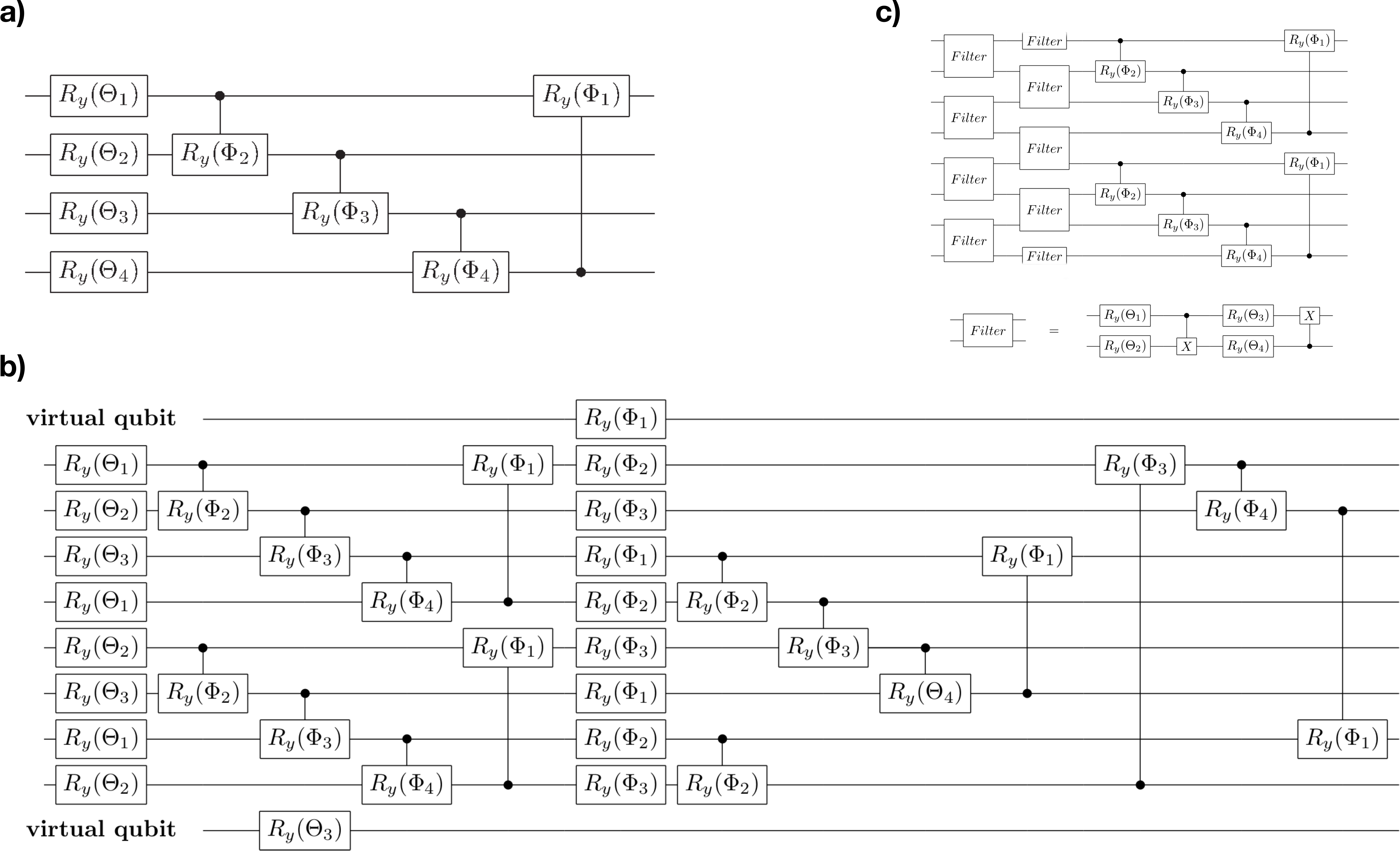}
        \vskip4mm
        \caption{Quantum circuits for preliminary scanning:
        In {\bf a)} the 4-qubit filter with 4-qubit entanglement shown;
	in {\bf b)} the stack of 3-qubit filters with 4-qubit entanglement is presented;
	in {\bf c)} the 2-qubit filters with 4-qubit entanglement} demonstrated.
        \label{fig:filtres}
\end{figure*}

The core concept that we use here is the hybrid (variational or quantum-classical) approach (for a review, see Refs.~\cite{Babbush2021-4,Aspuru-Guzik2022}).
This approach use parametrized (variational) quantum circuits, where the exact parameters of quantum gates within the circuit can be changed. 
The general structure of our variational circuit is represented on Fig.~\ref{fig:General}. 
Below we describe the proposed approach for multiclass classification based on the classical-quantum approach.

At the first step, we realize an amplitude encoding of input data, in our case, MNIST images. 
In fact, due to the high cost of this step~\cite{Biamonte2017}, we generate a set of encoding circuits, and store their parameters and structure in a memory, thus, hereby making the quantum version of dataset. 
We consider MNIST images, which are rescaled from 28 by 28 to 16 by 16 pixels, and, thus, 8 qubits are needed. 
In terms of the corresponding qubit states, encoded images can be expressed as follows:
\begin{equation}
	\Psi_{k} = \sum_{m=0}^{N}C_{m}^{k}|m\rangle,
\end{equation}
where $k$ is the index of image and $|m\rangle$ is a qubit register of 8 qubits, which encode index $m$, and $N=255$. 
Coefficients $C_{m}^{k}$ are equal to elements of normalized flatten vectors of images. 
In general, this approach enables us to pack vector of N double-precision numbers into log$_{2}$(N) qubits, and, thus, significantly reduce the size of processed data. 
It should be noted however that existing algorithms for amplitude encoding scales exponentially with N; further study is needed to overcome this problem.

We first employ the amplitude encoding procedure~\cite{Salomaa2005}, where ancilla qubits are used for one-hot encoding of the class of target images. 
Preliminary analysis of encoded images is performed with 3 convolutional layers with the sizes of filters, equal to $4$, $3$, and $2$, respectively. 
Each such layer consists of $2$ sublayers that are needed to maintain translational invariance (at least, partially),
and all the filters of the same size contain identical trainable parameters as it takes place the case for classical convolutional neural networks (CCNN). 
We note that for filters with the size of $3$ we need a virtual qubit, which is always set to zero; 
such a trick is needed to fit the filter into $8$ qubits in the translationally invariant manner. 
The convolutional layer with pooling is then placed after preliminary layers; at this step the first reduction of the required qubit number is realized. 

As in the classical setup, several fully connected layers are added after convolutional layers (9 layers in our case). 
The further reduction of qubit numbers is realized after regular layers and subsequent pooling are done (in the same way as it is done after convolutional filters). 

The final filter is needed for mixing the information from two parts of divided circuit. 
In the process of learning the output of final filter would contain the codes of classes: $|00\rangle$, $|01\rangle$, $|10\rangle$, and $|11\rangle$.
Output cascade contains four Toffoli gates, which activate the corresponding ancilla qubit; at the end of the quantum circuit we have one-hot encoded by ancillas class of image. 
Measurement results of ancilla qubits are passed to the \textit{softmax} activation function. 
The categorial cross-entropy is then used as the cost function. 
The subsequent calculations of gradients of the cost function with respect to the parameters of gates are done using parameter shift rule 
New parameters of quantum gates obtained by the gradient descent step.  
The detailed structure of all layers is described below. 

\section{Structures of layers}\label{sec:layers}

Here we present the detailed description of the layers that are used in our quantum machine learning algorithm. 

\subsection{Preliminary scanning using $n$-qubit filters}

The structure of 4-qubit filters is presented in Fig.~\ref{fig:filtres}a. 
First of all, $\RY(\Theta_{1})$, $\RY(\Theta_{2})$, $\RY(\Theta_{3})$ and  $\RY(\Theta_{4})$ rotations are added in order 
to rotate each of four qubit separately. 
We propose to use controlled parameterized rotations $\RY(\Phi)$ for the entanglement, which  is an essential new element in the structure of quantum perceptron.
We note that in Ref.~\cite{Chalumuri2021} authors use the standard controlled $\X$ gates for this purpose. 
Here, as we demonstrate, the parameterized entanglement scheme provide higher accuracy of image classification due to the more flexible learning algorithm. 

In classical machine learning, the linear perceptron is passed through a certain non-linear function, which is essential for the learning process. 
In the quantum case, instead of summations of neurons we use entanglement of qubits. 
The degree of entanglement is controlled by parameters $\Phi$, which makes learning process more flexible, and, thus, classification procedure may become more accurate. 
In fact, many classical activation functions like {\it sigmoid} or {\it tanh} behave akin to switches, so their values change from $0$ or from $-1$ to $1$ in a certain region. 
In quantum domain, we can switch from separable (non-entangled) to entangled state, what could play the role of non-linearities in classical learning. 
So far, individual rotations, which are followed by the parameterized entanglement, can be considered as an analog of the perceptron with the non-linearity. 

After 4-qubit scanning, smaller-scale filters are applied to analyze obtained quantum feature map in more details. 
The structure of layers with 3-qubit filters is presented in Fig.~\ref{fig:filtres}b. 
In order to rotate 3 individual qubits $\RY(\Theta_{1})$, $\RY(\Theta_{2})$ and $\RY(\Theta_{3})$ gates are added.
Similarly to the case of 4-qubit filters, individual rotations are performed by parameterized $\RY$ gates. 
We note that even in the case of 3-qubit filters, we use the entanglement of 4 qubits. 
Even though, the entanglement of 3 qubits looks more intuitive in this case, as we show below, the 4-qubit one provide more accurate results on image recognition. 
More detailed scanning of images is performed by layer with 2-qubit filters; the corresponding circuit is given in Fig.~\ref{fig:filtres}c (also see Ref.~\cite{Hur2022}). 
As in all previous cases, we use 4-qubit entanglement and the filter consist of 4 individual rotations with additional entanglement by \CNOT gates. 
The idea of using 4-qubit entanglement is inspired by classical CNN, where generation of new feature maps is done
by summation of contracted with weights previous feature maps and subsequent application of non-linearity.

\subsection{QCNN layer with pooling}

\begin{figure}[t]
        \centering
        \includegraphics[width=9.5cm]{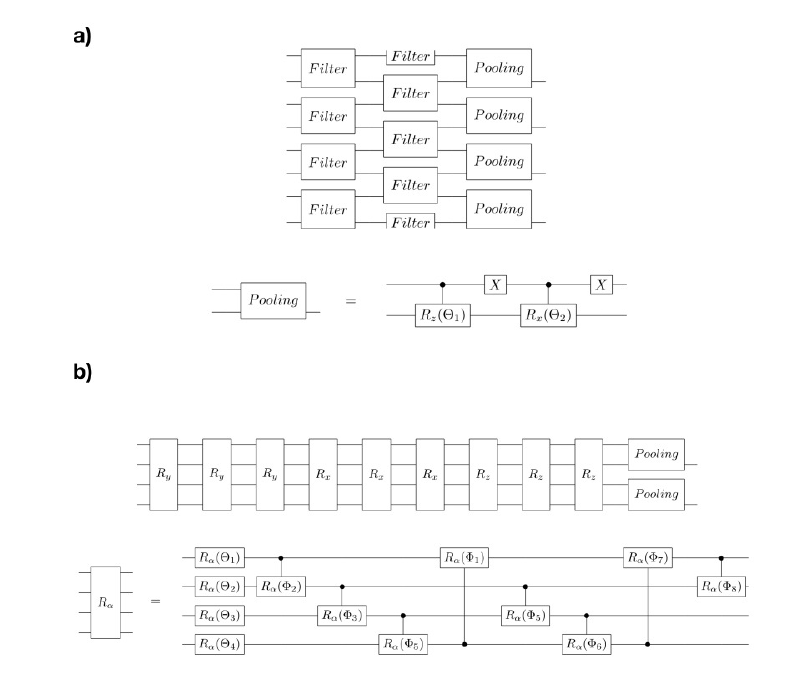}
        \vskip4mm
        \caption{In {\bf a)} the convolutional layer with pooling is shown. In {\bf b)} the structure of regular layers is illustrated.}
        \label{fig:layers}
\end{figure}

After the preliminary scanning step, the obtained quantum state of 8 qubits contains encoding of feature maps. 
The role of the next layer (see Fig.~\ref{fig:layers}) is to analyze these maps in more detail and pick up the most important of them. 
The scheme of the layer is given in Fig.~\ref{fig:layers}b, where the convolutional filter is the same as in Fig.~\ref{fig:layers}a. 
We note that in the pooling circuit, controlled $\RZ$ rotation is activated if the first qubit is at state $1$, while the controlled $\RX$ gate is used when upper qubit at state $0$.
This is conceptually similar to the structure proposed in Ref.~\cite{Hur2022}.

\subsection{Regular layers}

Similarly to the CCNN case, several regular layers are placed after convolutional layers. 
In our case we add 8 layers, as shown in Fig.~\ref{fig:layers}a. 
In order to get more accurate results, the double entanglement is added after individual rotations. 
The second reduction of qubit number in circuit is done by two pooling procedure as in the case of convolution layers. 
In order to obtain the required structure, we add a final filter at the end of the quantum circuit.  
As it is shown below, the use of the final filter is essential for obtaining more accurate results of image classifications.

\subsection{Toffoli and controlled rotation gates}

\begin{figure}[t]
        \centering
        \includegraphics[width=6cm]{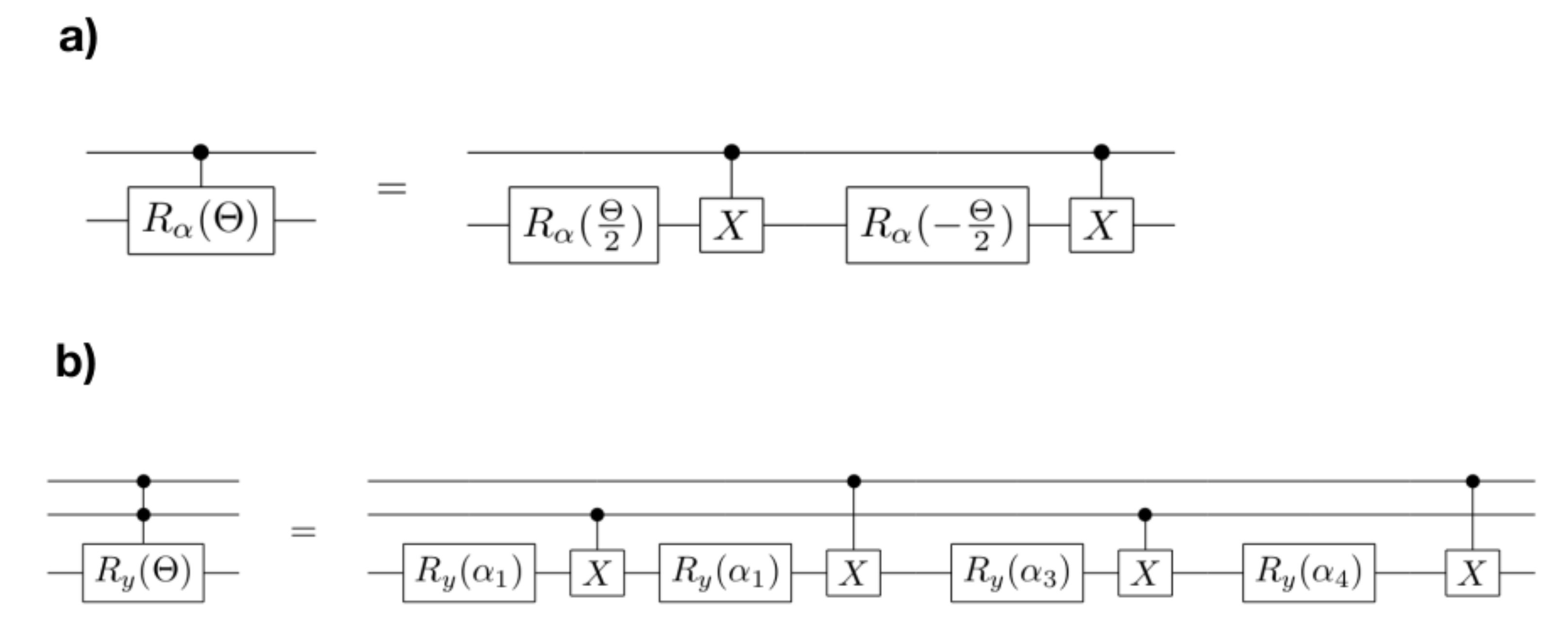}
        \vskip 5mm
        \caption{{\bf a)} Single-controlled rotation gates in terms of rotations and $\CNOT$ gates. {\bf b)} Decomposition of the Toffoli gate in terms of $\RY$ rotations and $\CNOT$ gates.}
        \label{fig:decomposition}
\end{figure}

The practical realization of high-fidelity two-qubit operations on quantum hardware is still a challenging task. 
The situation is typically more difficult with for three-qubit gates, such as Toffoli gate. 
Thus, it is necessary to decompose these gates via single- and two-qubit gates, which can be practically performed. 
The general algorithm of $n$-controlled rotations is presented in Ref.~\cite{Salomaa2004} and for the case of single-controlled rotation it can be expressed as it is shown in Fig.~\ref{fig:decomposition}a. 
In order to implement the Toffoli gate, we consider the qubit inversion as a rotation operation around $X$ or $Y$ axes and in our case doubly-controlled $\RY(\Theta)$ gate is used with the value of $\Theta=2\pi$. 
The circuit is presented in Fig.~\ref{fig:decomposition}b and it corresponds to the representation of sum of parameterized $n$-controlled rotations, which are considered in Ref.~\cite{Salomaa2004}. 
Toffoli gate, in fact, can be considered as a sum of such rotations with $n=2$, where $\Theta$ angles of all rotations, except the one that is controlled by 11th combination, are set to zero.
The definition of $\alpha$ angles is realized along the lines of the procedure of Ref.~\cite{Salomaa2004}; they are  obtained from  $\Theta$ angles by simple matrix transformation.

We note that multiqubit gate decomposition can be further improved using qudits, which are multilevel quantum systems.
As it has been shown, the upper levels of qudits can be used instead of ancilla qubits in the decomposition~\cite{Kiktenko2020,Kwek2021,Nikolaeva2021,Nikolaeva2021-2,Chong2019}.


\section{Classification results}\label{sec:results}

\begin{figure}[b]
	\centering
	\includegraphics[width=\columnwidth]{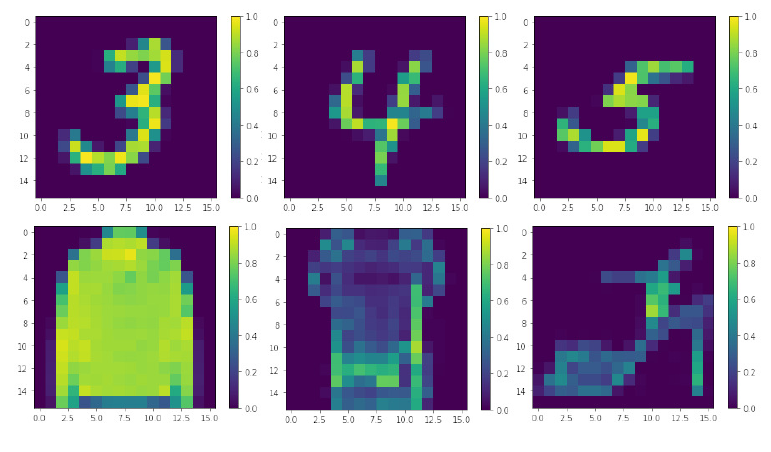}
	\caption{Examples of MNIST digits (top) and MNIST fashion (buttom) 
	images used in experiments.}
	\label{fig:MNIST}
\end{figure}

We benchmark the proposed quantum machine learning algorithm with the use of hand-written digits from MNIST and clothes images from fashion MNIST datasets. 
Examples are presented in Fig.~\ref{fig:MNIST}. 

All the simulations are performed using Cirq python library for the constructions of quantum circuits; 
TensorFlowQuantum library~\cite{Babbush2020} is used for the implementation of machine learning algorithm with parameterized quantum circuits. 
We use the Adam version of gradient descent with learning rate equal to $0.00005$, the overall number of trainable parameters in the QCNN circuit is equal to $149$. 
As a metric for the model performance, we simply use the accuracy of the recognition and for more detailed analysis two sets of experiments are done. 
In all the conducted experiments, parameterized quantum circuits are trained during 50 epochs.

Within the first set training and classification is done for the case, when dataset consists of images, which has certain similarity and, thus, classification problem become more difficult. 
We use MNIST images of digits 3, 4, 5, and 6 for this part. 
Also, fashion MNIST images with labels 0, 1, 2, and 3 are used for this purpose.

The second experimental set is focused on images, which are strongly differs from each other, thus, making recognition process easier; MNIST digits 0, 1, 2, and 3; 
fashion MNIST images with labels 1, 2, 8, and 9 are considered. 
Total number of considered images of each type is given in Table~\ref{bra:Tab1}. 

\begin{table}[h]
\begin{center}
\begin{tabular}{|c|c|c|c|c|c|c|c|}
\hline
MNIST digits & 0 & 1 & 2 & 3 & 4 & 5 & 6 \\
\hline
Training & 5923 & 6742 & 5958 & 6132 & 5842 & 5421 & 5918 \\
Test & 980 & 1135 & 1032 & 1010 & 982 & 892 & 958 \\
\hline
MNIST fashion & 0 & 1 & 2 & 3 & 8 & 9 &  \\
\hline
Training & 6000 & 6000 & 6000 & 6000 & 6000 & 6000 & \\
Test & 1000 & 1000 & 1000 & 1000 & 1000 & 1000 & \\
\hline
\end{tabular}
\caption{Number of images of each type.}
\label{bra:Tab1}
\end{center}
\end{table}

Each image vector is normalized to one since only such kind of vectors can be used by amplitude encoding algorithm. 
Results of image classification are given in Table \ref{bra:Tab2}. 
Quantum circuits for multiclass classification are considered in Ref ~\cite{Chalumuri2021}. 
QCNN examples, provided within documentation of TensorFlowQuantum~\cite{Babbush2020} also can be relatively simply generalized for the case of multiclass classification tasks. 
In the second column of Table \ref{bra:Tab2}, we provide results of experiments with circuits, similar to those of Ref~\cite{Chalumuri2021}. 
In order to obtain these results we replace all the $\RY(\Theta)$ used at entanglement steps by CNOT gates. 
Also, we remove all the parts of circuit of Fig.~\ref{fig:General}, which are placed after regular layers, i.e. pooling layers, final layers and the part with Toffoli gates. 
Entanglement of ancilla qubits with regular layers is done via CNOT gates according to Fig. 2 of Ref~\cite{Chalumuri2021}. 
Third column of Table~\ref{bra:Tab2} contains results, which are obtained with the full circuit of Fig.~\ref{fig:General}. 
Significant improvement of the accuracy of classification results caused by two facts. First one is the usage of parameterized entanglement in our circuit. 
Secondly, an increase of the performance may be connected with the fact, that our circuit constructed in a similar way to classical neural networks, 
we use qubit reduction procedure in analogy with the reduction of number of layers outputs in classical case until the number of outputs become equal to the number of target classes. 
Note that in Fig. ~\ref{fig:General} ancilla qubits are used only at read-out step and no entanglement is needed between ancillas and other qubits during computational procedure, 
what can significantly simplify requirements to the corresponding quantum hardware.
We also compare obtained quantum results with results of the CCNN with similar number of parameters, which is 188 in our case. 
The structure of the CCNN is presented in Table \ref{bra:Tab3}.
Clearly, classical results are more accurate, what indicate on the fact that with similar number of parameters classical model is still more expressive. 
An analysis of possible quantum advantage in ML tasks is presented in Ref~\cite{Boxio}. 
In their study authors analyze ML models based on kernel functions and show that with enough data provided classical methods become more powerful then corresponding quantum algorithms. 
Thus, additional study is still needed to find ML tasks where quantum algorithms will outperform their classical analogs.

\begin{table}[h]
\begin{center}
\begin{tabular}{|c|c|c|c|}
\hline
& Quantum & Quantum & Classical \\
& ref.~\cite{Chalumuri2021} - like & Fig.~\ref{fig:General} & \\
\hline
MNIST digits (3456) & 71.44 & 85.14 & 94.25 \\
MNIST digits (0123) & 77.64 & 90.03 & 95.85 \\
MNIST fashion (0123) & 71.15 & 85.93 & 89.69 \\
MNIST fashion (1289) & 79.33 & 93.63 & 97.42 \\
\hline 
\end{tabular}
\caption{Accuracies of classification for quantum and classical convolutional neural networks.}
\label{bra:Tab2}
\end{center}
\end{table}

\begin{table}[h]
\begin{center}
\begin{tabular}{|c|c|c|}
\hline
Layer type & Output shape & Number of parameters \\
\hline
Conv2D &  (None, 14, 14, 1) & 10 \\
Conv2D &  (None, 12, 12, 1) & 10 \\
Pooling & (None, 6, 6, 1)  & 0 \\
flatten & (None, 36)  & 0 \\
Dense & (None, 4) & 148 \\
Dense & (None, 4) & 20 \\
\hline 
\end{tabular}
\caption{Structure of the used classical convolutional neural network with 188 parameters.}
\label{bra:Tab3}
\end{center}
\end{table}

In overall, the QCNN can produce accuracy of multiclass classification that are qualitatively similar to the classical model if the number of parameters are comparable.
We would like to mention that the similar level of the accuracy has been achieved in Ref.~\cite{Chalumuri2021} for the case of the 3-class classification problem. 
Here we have demonstrated this level of the accuracy for the 4-class classification tasks, which to the best of our knowledge is the first such demonstration.

\section{Conclusion}\label{sec:conclusion}

Here we have demonstrated the quantum multiclass classifier, which is based on the QCNN architecture. 
The main conceptual improvements that we have realized are the new model for quantum perceptron and the optimized structure of the quantum circuit.
We have shown the use of the proposed approach for 4-class classification for the case of four MNIST.
As we have presented, the results obtained with the QCNN are comparable with those of CCNN for the case if the number of parameters are comparable.
We expect that further optimizations of the perceptron can be studied in the future in order to make this approach more efficient. 
Moreover, since the scheme require the use of multiqubit gates, the qudit processors, 
where multiqubit gate decompositions can be implemented in a more efficient manner, can be of interest for the realization of such algorithms. 

\newpage

\section*{Acknowledgments}
We thank A. Gircha for useful comments.
A.S.M. and A.K.F. acknowledge the support of the Russian Science Foundation (19-71-10092).
This work was also supported by the Russian Roadmap on Quantum Computing.

\bibliography{bibliography.bib}

\end{document}